# The Filter Wheel and Filters development for the X-IFU instruments on-board Athena

E. Bozzo[*a], M. Barbera[b], L. Genolet[a], S. Paltani[a], M. Sordet[a], G. Branduardi-Raymont[c], G. Rauw[d], S. Sciortino[e], D. Barret[f,g], J. W. Den Herder[h]

[a]Department of Astronomy, University of Geneva, Chemin d'Ecogia 16, 1290, Versoix, Switzerland;
[b]UNIPA/Dipartimento di Fisica e Chimica, Palermo, Italy;
[c]UCL, Department of Space and Climate Physics, MSSL, Holmbury St Mary, Dorking, Surrey RH5 6NT UK;
[d]Université de Liège, Département Astrophysique, Géophysique & Océanographie, Allée du 6 Aout 19c, Bat. B5c 4000 Liege Belgium;
[e]INAF/Osservatorio Astronomico di Palermo, Piazza del Parlamento 1, 90134 Palermo, Italy;
[f]Universite de Toulouse, UPS-OMP, IRAP, Toulouse, France;
[g]CNRS, IRAP, 9 Av. colonel Roche, BP 44346, F-31028 Toulouse cedex 4, France;
[h]SRON, Netherlands Institute for Space Research, Utrecht, The Netherlands.

## ABSTRACT

Athena is the large mission selected by ESA in 2013 to investigate the science theme "Hot and Energetic Universe" and presently scheduled for launch in 2028. One of the two instruments located at the focus of the 12 m-long Athena telescope is the X-ray Integral Field Unit (X-IFU). This is an array of TES micro-calorimeters that will be operated at temperatures of 50 mK in order to perform high resolution spectroscopy with an energy resolution down to 2.5 eV at energies < 7 keV. In order to cope with the large dynamical range of X-ray fluxes spanned by the celestial objects Athena will be observing, the X-IFU will be equipped with a filter wheel. This will allow the user to fine tune the instrument set-up based on the nature of the target, thus optimizing the scientific outcomes of the observation. A few positions of the filter wheel will also be used to host a calibration source and to allow the measurement of the instrument intrinsic background.

## 1. INTRODUCTION

Athena [1] is the future X-ray observatory designed to address the "Hot and Energetic Universe" science theme and selected in 2013 as the L2 mission by the European Space Agency (ESA) within the Cosmic Vision science program. At present, the expected launch date of Athena is in 2028. The mission is equipped with a 12-m long telescope, featuring Silicon Pore Optics (SPOs, [3]), which achieve an unprecedentedly large collecting area for X-ray photons of ~2 m$^2$ at 1 keV. At the focus of the Athena telescope there will be two instruments: the Wide Field Imager (WFI, [4]) and the X-ray Integral Field Unit (X-IFU, [2]). The latter is an array of transition edge sensor (TES) micro-calorimeters operating at a temperature of 50 mK and providing simultaneously a high spectral resolution ($\Delta E_{FWHM}$ = 2.5 eV at E < 7 keV) and imaging capabilities in the energy range 0.2-12 keV. The X-IFU will have a field of view with a diameter of 5' and will be equipped with a filter wheel (FW). The different filters on the FW allow the scientific user to tailor the configuration of the instrument to the target source in order to maximize the scientific output of the observation. Dedicated FW positions will be used to host a calibration source and a closed position to allow the estimation of the instrument intrinsic background and to protect the instrument when not in use. In this contribution, we describe the current status of the X-

---

[*]enrico.bozzo@unige.ch

IFU FW development at the University of Geneva (UoG, Switzerland) and the on-going trade-offs for the filters being carried out in collaboration with the University of Palermo (Italy).

## 2. THE HERITAGE OF THE UNIVERSITY OF GENEVA

The X-IFU FW is being designed by the University of Geneva (UoG) based on the heritage gained with the Astro-H mission ([5]). We developed the FW mechanism (FWM) for Astro-H, which has been mounted about 920 mm above the Soft X-ray Spectrometer (SXS), and its associated commanding electronics (FEW). The main function of the FW is to provide the capability to move different filters into the X-ray beam, before the latter reaches the instrument focal plane. The flight model (FM) of the SXS FW was successfully delivered by the UoG to the Japanese space agency (JAXA) at the end of 2013 and launched on-board Astro-H in 2016.

The FM of the SXS FW is shown in Figure 2-1. The photo on the left side shows the bottom of the filter wheel housing with a dummy filter holder. The photo on the right side shows a large opened cylinder on which the modulated X-ray sources (MXSs, [6]) are mounted, two rectangular boxes which are the high voltage power supplies (HVPSs) needed for the functioning of the MXSs and the top of the housing. The filter wheel hosting six positions (filters not yet in place) is mounted on a shaft held by wet lubricated angular ball bearings. The FWM is actuated by a stepper motor located in an eccentric position with respect to the FWM rotation axis. A planetary gear is embedded in the stepper motor and connected to the FWM with a pinion and a spur gear. A potentiometer is used to monitor the FWM position. The placement of all FW components mentioned so far is shown in Figure 2-2. The MXSs, which are used for the calibration of the SXS instrument, are mounted on the FWM housing. In the Astro-H project, both the MXSs and the corresponding HVPS (also mounted on the FW housing) were developed and provided under the responsibility of SRON (The Netherlands) but controlled by the FWE that was part of the UoG contribution.

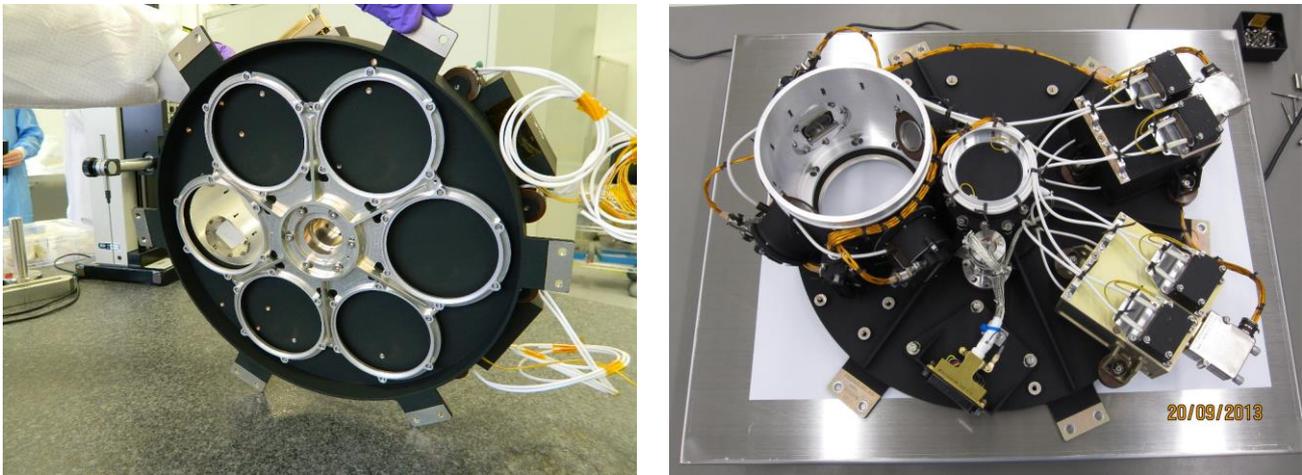

**Figure 2-1 The Astro-H FW developed by the University of Geneva and the Swiss industry Ruag Space. The approximate size of the overall FWM shown above is 351 mm (length) x 384mm (width) x 100mm (height). The estimated mass, including the harness and the HVPSs is of 2.5 Kg. The figure on the left shows the six filter positions and the central hole where the shaft, regulating the FWM rotation, has to be placed. The figure on the right is taken from the opposite side and shows the details of the FWM housing hosting the opened cylinder in which the MXSs are located, the motor, and the two HVPS with the corresponding cables. The connector that is visible on the FWM housing connects the FWM with the FWE.**

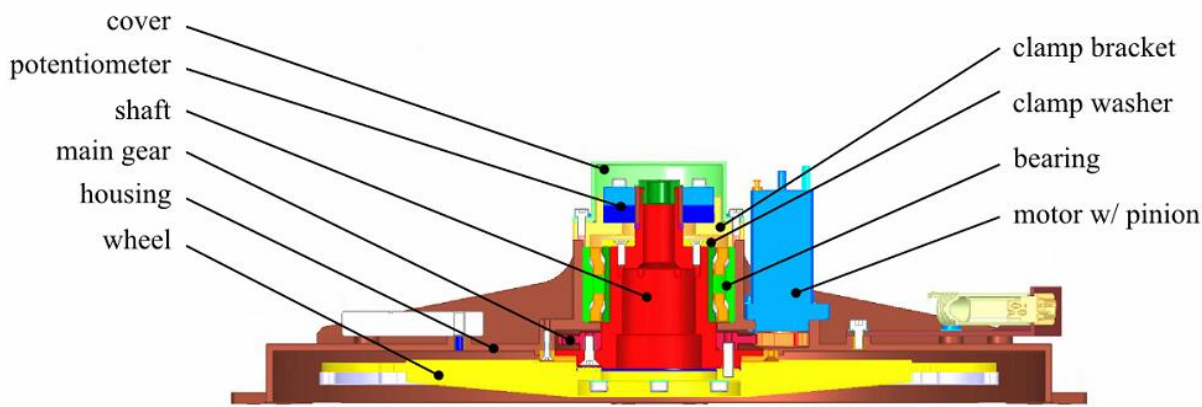

**Figure 2-2 The Astro-H FWM cross-section showing the location of the most relevant parts and of the components of the mechanism.**

The details of the FWE developed for the Astro-H FW are shown in Figure 2-3. The FWE is composed of an aluminum mechanical box housing two electronic boards (the data board and the power board) and two DC-DC converters. The FWE architecture block diagram is represented in the top left side of Figure 2-3, while the functional diagram is shown on the top right side. The FWE main functions are:

- To communicate with the spacecraft through redundant space-wire link (10Mbits/s; in Astro-H a RMAP protocol was adopted).
- To generate the voltages required by the FWM stepper motor, the MXS calibration sources, and the FWE itself from the power provided by the spacecraft Power Control Unit (PCU). In Astro-H the PCU was providing an input power of 32.1 to 52.5V and the linear regulators of the FWE were able to provide the following voltages from the DC-DC converters output voltages:
    - +1.5V and 3.3V for the field programmable gate array (FPGA);
    - +3.3V for the house-keeping analog-to-digital converter;
    - +2.5V and +5V voltage reference for the house-keeping parameters measurements;
    - +10.1V for the four MXS high voltage control;
    - +12.1V and -12.9V to power the FW stepper motor;
    - +12.4 and -13.2V for the analog electronic part.

  Current pulses of 200mA (maximal current) and 30mA (average current) were also produced by the linear regulators in order to power the four LEDs of the MXSs. An overvoltage protection was connected on the +14.5V to protect the FWE against possible failures of the DC-DC converters.
- To drive the filter wheel motor and to read its position using a potentiometer.
- To switch on and off the electronically controlled MXS calibration sources and the corresponding HVPSs.
- To acquire and broadcast house-keeping parameter data (temperatures, voltages, currents...).

The FWE of Astro-H was capable to independently and simultaneously operate the stepper motor, the position encoder, and the calibration sources. To guarantee the full redundancy of the unit, the stepper motor was also equipped with two independent windings that can be powered in parallel to boost the FW rotation in case of emergency. The reliability analysis carried out during the development of the FWE showed that a design with a single FPGA was preferable compared to a fully redundant design with two FPGAs in order to avoid any cross-strapping of the electrical connections. In Astro-H both the FWM and the FWE were operating at ambient temperature and no operations in cold were foreseen.

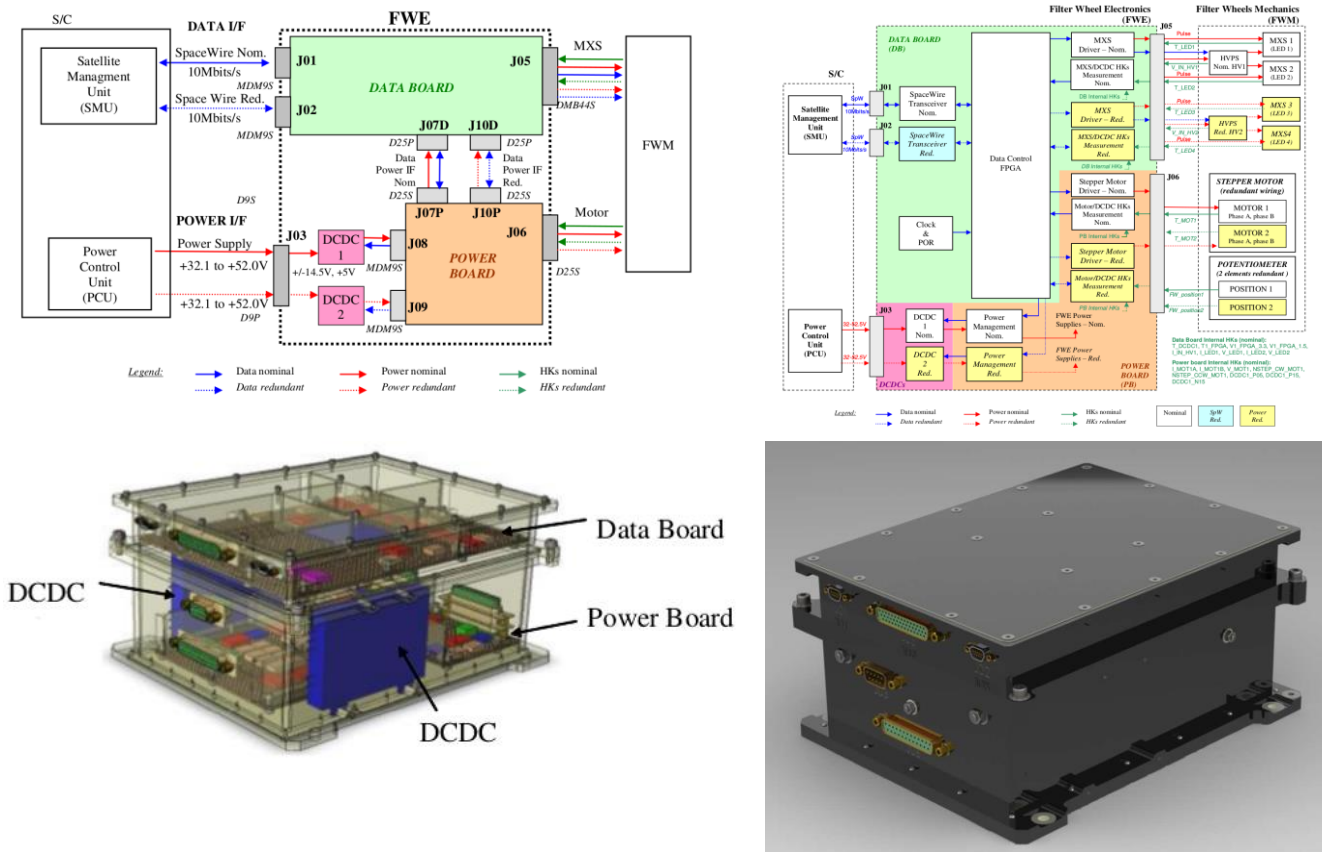

**Figure 2-3** Top left: the FWE architecture block diagram, showing the connection of the FWE to the Astro-H spacecraft, as well as to the FWM. Both the data and the power boards are indicated, together with the two DC-DC converters. Top right: the FWE functional diagram. Bottom left and right: a schematic representation of the FWE showing the appearance of the unit flight model and the internal location of the different components. The mechanical envelop of the FM of the FWE in Astro-H was about 101x201x170 mm$^3$.

## 3. THE X-IFU FILTER WHEEL CONCEPT

The current baseline for the Athena X-IFU FW and FWE foresees an electrical architecture diagram and a mechanical structure relatively similar to those discussed in Sect. 2, but in a direct-drive configuration (i.e. with the motor centrally placed on the FWM). A number of additional modifications are already foreseen with respect to the Astro-H case in order to comply with the more stringent requirements provided by ESA for the X-IFU instrument.

For what concerns the FWM, the major change involves the increase of the FWM size to include up to 8 different filter positions. The aim is to provide through this improved FWM design:
1. Protection of the X-IFU focal plane detectors against micro meteorites and contamination;
2. Reduction of the optical load from bright stars which could degrade the energy resolution of the instrument;
3. Possibility to optimize the observational throughput in case of bright X-ray targets;
4. Capability to separate the sky contributions from the intrinsic instrumental background;
5. Provision of a radioactive calibration source as a backup to the MXSs.

Although a number of trade-off studies are still on-going to confirm the types of filters needed to fulfill the X-IFU scientific objectives, the plan is to provide the above functionalities by including among the filter wheel positions:

1. One closed position (Molybdenum filter), to be used during launch for the protection of the focal plane detectors and in-flight for the evaluation of the internal detectors background (the filter will screen out all X-ray photons coming from the telescope beam).
2. One open position (no filter) that can be used for the observations of faint X-ray sources where no contamination from bright optical targets is expected or in the event of failures related to the FW.
3. Two Beryllium filters of different thickness that are able to suppress to different extents the X-ray fluxes of celestial sources at energies < 3 keV, where the bulk of the X-ray photons are usually emitted. These two filters can thus be adopted to limit the degradation of the X-IFU performances during the observations of particularly soft and bright X-ray sources that are characterized by fluxes as high as ~50 mCrab (this value is still being assessed by proper simulations).
4. A single neutral density blocking filter that can be used to induce an overall decrease of the rate of X-ray photons coming from particularly bright sources onto the X-IFU focal plane by a factor that could be as high as 100. The latter number is still being evaluated depending on the final scientific goals of the X-IFU and could be improved in order to allow the observations of X-ray sources with fluxes of 100 mCrab or brighter.
5. Two optical blocking filters that are needed to limit the optical load (and thus the degradation of the instrument energy resolution) from the bright UV/V counterparts of the X-ray sources to be observed by the X-IFU. We specifically address the need of these filters in Sect. 0.

In addition to the filters mentioned above, the X-IFU FWE will also drive the MXSs, as it was in the case of Astro-H. Even though we do not enter the details of the MXS development in this contribution, it is worth noticing that a similar design to that of the Astro-H MXSs is being presently adopted as a baseline for the X-IFU.

The preliminarily sketch of the X-IFU filter wheel filter implementation is shown in Figure 3-1. The introduction of 8 instead of 6 filter positions (as it was in Astro-H) leads to a substantial increase in the lateral size of the FWM (as indicated in the drawings compared to the measurements reported in the caption of Figure 2-1; the updated vertical size still needs to be evaluated). The impact on the total mass of the device is still being assessed, as well as the corresponding consequences for the exported torque and micro-vibrations induced by the device in operations. Mass optimization trade-offs are on-going, considering also improved solutions for the holding structures of the filters and the housing of the stepper motor.

The FWE for the X-IFU FW is being currently designed to be similar to the Astro-H case depicted in Sect. 2. The main trade-off being investigated is the possibility of including a second fully redundant FPGA, in order to comply with the redundancy requirements imposed by ESA for all Athena instrument sub-systems. A reliability analysis is being carried out to assess the impact of the cross-strapping of the electrical connections that could be used to implement the two FPGAs in the FEW data board.

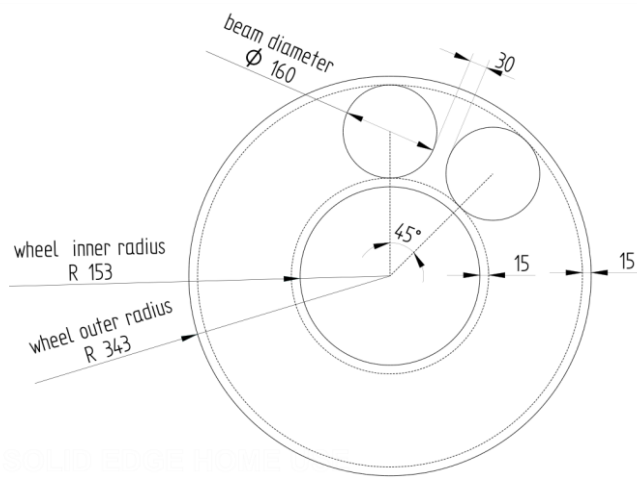

**Figure 3-1** A preliminarily evaluation of the increase in the lateral size of the Athena X-IFU filter wheel when 8 filters instead of 6 (as it was in Astro-H) are included. The increase can be appreciated by comparing the numbers reported here with those indicated in Figure 2-1.

# 4. X-IFU OPTICAL FILTERS

A number of studies are being performed to optimize the design of the X-IFU FW filters. In this section we specifically address the need of optical filters to fulfill some of the X-IFU science goals, in particular that of maintaining the highest level of the instrument spectral resolution while maximizing its sensitivity to soft X-rays (below ~0.5 keV). These filters are being intensively studied by the X-IFU team at the University of Palermo (UoP).

The optical load from the bright UV/V counterparts of some of the X-ray sources that will be observed with the X-IFU in order to fulfill the Athena scientific goals (e.g. massive O-B hot stars as ζ Pup; bright late-type stars as Altair and Sirius A+B; planets as Mars, that is the most challenging target in this category having 3.8 mag/arcsec$^2$) can degrade the X-IFU energy resolution by significantly increasing the so-called photon shot noise (i.e. the noise related to the Poissonian fluctuations in the number of photons hitting the detectors on the X-IFU focal plane). To evaluate the optical load on the X-IFU instrument in a conservatively worst case, we simulated a number of Athena observations of massive hot stars. A requirement has been recently established in order to secure the observability of these objects with the X-IFU up to magnitudes V=2 (e.g. ζ Pup, ζ Ori, and ζ Oph) and probe the dynamics of massive star winds with high resolution X-ray spectroscopic measurements ([7]). The spectra of massive stars were characterized in our simulations by using synthetic data produced from the Ostar2002 and Bstar2006 grids adopting the non-LTE plane-parallel model atmosphere code TLUSTY ([8]). The output files from the grids provide the Eddington fluxes at the stellar surface as a function of frequency from the EUV into the far IR. The Eddington fluxes were then converted into fluxes at Earth for a given spectral type and a given apparent magnitude. Bolometric corrections, needed for this conversion, were taken from [9] for the O-type stars and [10] for the B-type stars. Only main-sequence stars were considered, as the relevant fluxes depend mainly on effective temperature and apparent magnitude and are not very sensitive to the surface gravity. We considered a total of 22 temperatures between 15000 and 45000 K and simulated 20 different apparent V-band magnitudes for each temperature, ranging from 2.0 to 11.5. Interstellar dust absorption was accounted for via the relation provided by [11]. We considered four values of the extinction: AV = 0.0, 0.5, 1.0, and 1.5. This reddening law does not deal with the absorption by the interstellar medium (ISM), which is especially important at wavelengths shorter than 912 Å where interstellar hydrogen blocks our view in most directions. We have thus assumed that no flux at wavelengths shorter than 912 Å enters the Athena telescope. This assumption breaks down for a few early-type stars that are located in known holes of the ISM. To comply with our requirement of simulating a worst case scenario, we have assumed AV = 0.0 which is a realistically conservative value for the optically brightest massive stars. The ISM absorbed fluxes at Earth were integrated between 912 and 12000 Å. For the simulations, we assumed a thermal filters configuration presently under study consisting of five equal filters, each one consisting of 45 nm of Polyimide and 30 nm of Aluminum ([13]). Furthermore, we assumed an effective collecting area for X-ray photons of 2 m$^2$, a fraction of 0.43 of the number of photons of a point source falling within the central pixel of the PSF, and an X-IFU detector read-out time of 7.5 ms corresponding to 50 times the nominal detector decay time of 150 us which is considered the record length necessary to achieve the maximum energy resolution. The energy resolution degradation of the X-IFU instrument can be computed as ([12]):

$$\Delta E_{FWHM} = 2.35 * \sqrt{dt * NEP^2(T)},$$

where:

$$NEP^2 = \int 2\, power\,(v)\, hv\, dv$$

and power is the power per unit frequency received by a single pixel.. We indicated above with *dt* the detector read-out time. The goal is to maintain through the usage of properly designed filters a noise contribution to the overall instrument energy resolution of <0.2 eV.

The results of the above computation are shown on the left side of Figure 4-1 with solid lines (different colors are used for the different temperatures of the considered stars). In these cases, only the thermal filters are assumed to be placed between the optical beam penetrating through the X-IFU dewar aperture cylinder ([2]) and the instrument focal plane detectors. As it can easily be noticed, even the faintest of these hot stars would have a non-negligible degradation of the instrument spectral resolution, thus impacting the scientific investigation of the stellar wind dynamics. In the same figure, we show with dashed lines the effect produced by the introduction of different optional optical blocking filters (OBF) in the X-IFU FW. We specifically analyzed the case of a "thin" OBF, consisting of 200 nm polyimide + 40 nm of Al, and a "thick" OBF, consisting of 200 nm polyimide + 80 nm Al. The thin OBF allows to observe hot stars fainter than mV ~ 6 with no significant degradation of energy resolution, while with the thick filter observations at the highest spectral resolution offered by the instrument can be carried out up to objects as bright as mV = 2. The X-ray transmission properties of the two filters in the X-IFU energy band are shown on the right side of Figure 4-1.

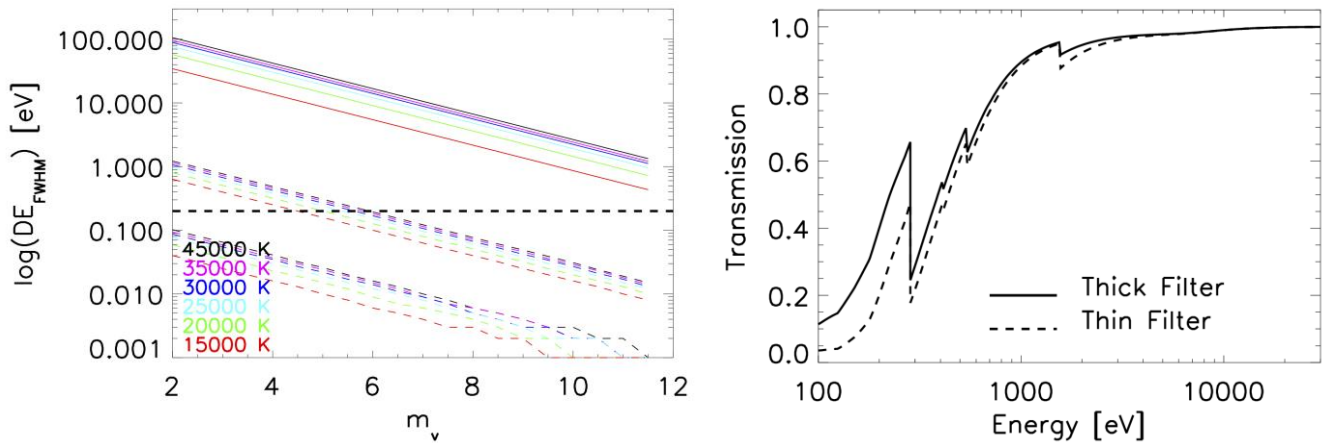

**Figure 4-1 Left: X-IFU energy resolution degradation due to optical load from hot O-B stars. The solid lines include UV/V stellar spectra attenuation by the thermal filters placed at the entrance of the X-IFU aperture cylinder, while the dashed lines include the presence of the additional optical filters mounted on the X-IFU FW. Right: the X-ray transmission of the optical blocking thin and thick filters considered in order to maintain the X-IFU energy resolution**

## 5. CONCLUSIONS

The UoG is currently in charge of designing and developing the FWM and FWE for the X-IFU instrument, based mainly on the heritage gained thanks to a similar unit provided for the Astro-H mission. The more stringent requirements and more challenging scientific goals of the X-IFU instrument require, however, a significant improvement of the Astro-H design. The main design activities are now focused especially on the increase in the FW size to include up to 8 filter positions, while maintaining the mass and power budget strictly under control. Trade-off on the filter types to be mounted in the X-IFU FW are also on-going. X-ray specific filters (e.g., Beryllium or gray filters) are need to optimize the observations of bright sources at high spectral resolution. Optical blocking filters are needed to cope with the large photon shot noise and induced energy degradation of the bright optical counterparts of some of the potential X-IFU science targets.

## ACKNOWLEDGEMENT

The Athena team at the University of Geneva acknowledges the support of the Swiss State Secretariat for Education, Research and Innovation SERI and ESA's PRODEX programme. UoG acknowledges support for the Astro-H FWM and FWE development from MCSE (Switzerland) and RUAG (Switzerland). S. Sciortino and M. Barbera acknowledge support by ASI (Italian Space Agency) through the Contract n. 2015-046-R.0.